\documentclass[fleqn,twoside]{article}
\usepackage{epsfig,espcrc2}

\usepackage{graphicx}
\usepackage[figuresright]{rotating}

\newcommand{\AmS}{{\protect\the\textfont2
  A\kern-.1667em\lower.5ex\hbox{M}\kern-.125emS}}

\title{Improving the Quark Number Susceptibilities for Staggered
Fermions
\vskip-1.8cm\hfill\small hep-lat/0209008, TIFR/TH/02-29\vskip1.5cm
}

\author{Rajiv V. Gavai\address{Department of Theoretical Physics, Tata
                Institute of Fundamental Research, \\
                Homi Bhabha Road, Mumbai 400 005, India}%
        \thanks{E-mail: gavai@tifr.res.in} }
       
\begin{document}

\begin{abstract}

Quark number susceptibilities approach their ideal gas limit at sufficiently
high temperatures.  As in the case of other thermodynamic quantities, this
limit itself is altered substantially on lattices with small temporal extent,
$N_t$ = 4--8, making it thus difficult to check the validity of perturbation
theory.  Unlike other observables, improving susceptibilities or number
densities is subject to constraints of current conservation and absence of
chemical potential ($\mu$) dependent divergences.  We construct such an
improved number density and susceptibility for staggered fermions and show that
they approximate the continuum ideal gas limit better on small temporal
lattices.  

\vspace{1pc}
\end{abstract}

\maketitle

\section{INTRODUCTION}

Thermodynamic observables, such as the energy density $\epsilon$, the pressure
$P$ or the quark number susceptibility $\chi$, approach the corresponding ideal
gas values as the temperature $T$ (or the chemical potential $\mu$) becomes
very large. Corrections to the ideal gas behaviour can be computed in a weak
coupling approximation.  Such a picture of ideal or weakly interacting plasma
is often used in most physical applications in heavy ion phenomenology and in
cosmology of the Early Universe.  Whether the quark-gluon plasma can be
described by such a simple picture in the relevant range of temperatures, say
$1 \le T/T_c \le 10$, is a very important issue for these applications.
$\epsilon$ and $P$ are known to show strong deviations from the ideal gas
picture in this range which cannot be explained by naive perturbation theory up
to $O(g^5)$.  Clever resummations of the perturbation series are thought to be
a cure.  One such approach \cite{bir} successfully describes the lattice data
for $P$ above $\sim 3 T_c$.  The quark number susceptibilities provide an
independent test of any such resummation scheme. Recent lattice results
\cite{us} for these exhibit a strong disagreement with both the naive
perturbation theory as well as the resummed \cite{bir2} one.  These lattice
results were obtained using staggered fermions on lattices with 4, 6 and 8
temporal sites in a temperature range $1 \le T/T_c \le 3$.  The ratio
$\chi/\chi_{ideal}$ from these  calculations were extrapolated to the continuum
to compare with the perturbative results.

\begin{figure}[htbp]\begin{center}
\vspace{-0.7cm}
\epsfig{height=7.5cm,width=6cm,file=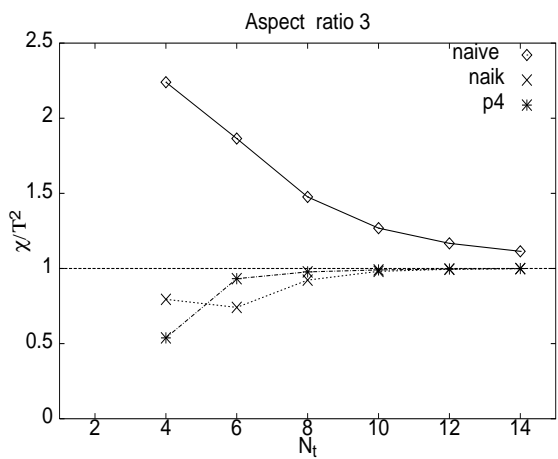,angle=270}
\vspace{-0.7cm}
\caption{ The susceptibility $\chi/T^2$ as a function of $N_t$ for
$N_s^3 \times N_t $ lattices with $N_s = 3 N_t$. }
\vspace{-0.7cm}
\label{chiT}\end{center}\end{figure}
\vspace{-0.3cm}

As shown in Figure \ref{chiT}, the ideal gas result, $\chi_{ideal}$, for the
naive staggered fermions varies strongly  for $N_t$ = 4, 6 and 8, and deviates
substantially from the continuum result, indicating the need for an improved
operator for it in order to confirm the implied non-perturbative nature of the
quark-gluon plasma by the results of \cite{us}.  Recall that $\chi$ (or the
number density) is obtained from $\ln Z$ by taking second (or first) derivative
with respect to $\mu$.  Improving it therefore needs to be done by modifying
the number operator $N$, which in turn is fixed by the (baryon) current
conservation equation.  Following the study \cite{urs} of improvement of the
energy and pressure, we employ the Naik action \cite{naik} and the P4 action
\cite{urs} and write down $N$ for them from the corresponding current
conservation equations.  An additional complication for nonzero $\mu$ is the
presence of divergences even for the ideal gas. Prescriptions \cite{us1} were
proposed for the naive action to eliminate them, which were shown \cite{us2} to
satisfy a general condition.  We demonstrate that the same condition can
suffice for a class of improved actions as well.  Numerical results for the
ideal gas are presented to contrast  $\chi(0,T)$, $n(0,T)$ and $n(\mu,T)$ with
the corresponding continuum results.

\section{IMPROVED NUMBER DENSITY}

We follow the notation of Ref. \cite{urs} for the general form of the 
fermion actions, which yields the naive, the Naik and the P4 action
for their $c_{1,0}$, $c_{3,0}$ and $c_{1,2}$ equal to ($\frac{1}{2}$,0,0),
($\frac{9}{16}$,$\frac{-1}{48}$, 0) and ($\frac{3}{8}$, 0, $\frac{1}{48}$)
respectively, where these coefficients multiply the first, third and mixed
third derivatives respectively. Setting all the link variables to unity, the
current conservation equation for the free theory can be easily shown to be :
\begin{eqnarray}
& \sum_{\mu} \bigg[ c_{1,0} \Delta_\mu j^{(1)}_\mu   +
c_{3,0} \Delta_{3\mu} j^{(3)}_\mu  + & \nonumber \\  
&c_{1,2}  \sum_{\nu \ne \mu} \bigg\{ \Delta_{\mu +2\nu} j^{(1)}_{\mu
\nu} + \Delta_{\mu - 2 \nu} j^{(2)}_{\mu \nu}  \bigg\} \bigg] = 0,~~&
\label{cc}
\end{eqnarray}
where the currents are defined by
\begin{eqnarray} 
j^{(1)}_\mu(x) =& \bar \psi_{x} \gamma_\mu \psi_{x+ \hat \mu} + \bar \psi_{x+
\hat \mu} \gamma_\mu \psi_{x},~~~~~~~~~~& \nonumber \\ 
j^{(3)}_\mu(x) =& \bar \psi_{x} \gamma_\mu \psi_{x+ 3 \hat \mu} + \bar \psi_{x+
3 \hat \mu} \gamma_\mu \psi_{x},~~~~~~~~~& \nonumber \\ 
j^{(1)}_{\mu \nu}(x)=& \bar \psi_{x} \gamma_\mu \psi_{x+ \hat \mu +2 \hat \nu}
+ \bar \psi_{x+ \hat \mu + 2 \hat \nu} \gamma_\mu \psi_{x},& \nonumber \\  
j^{(2)}_{\mu \nu}(x) =& \bar \psi_{x} \gamma_\mu \psi_{x+ \hat \mu -2 \hat \nu}
+ \bar \psi_{x+ \hat \mu - 2 \hat \nu} \gamma_\mu \psi_{x}.& \label{cur}
\end{eqnarray}
The $\Delta$'s in eq. (\ref{cc}) denote appropriate backward difference
operators.  Summing eq. (\ref{cc}) over all spatial lattice sites $\vec x$,
one obtains $\sum_{\vec x} N(x)$ = constant, where the ``number density''
\begin{eqnarray}
&N(x) = c_{1,0} j^{(1)}_0(x) + 3 c_{3,0} j^{(3)}_0(x) + &\nonumber \\
&2c_{1,2} \sum_{\mu \ne 0}[j^{(1)}_{\mu 0}(x) -  j^{(2)}_{\mu 0}(x)] + 
&\nonumber \\ 
&c_{1,2}\sum_{\mu \ne 0}[j^{(1)}_{0 \mu}(x) 
+ j^{(2)}_{0 \mu}(x)] + \Delta_t F(x)~.&
\label{baryon}
\end{eqnarray}
Adding the term $\mu \sum_{x} N(x)$ to the $\mu =0$ action, the last term drops
out due to the anti-periodic boundary conditions on the $\psi$ and $\bar \psi$
in the time direction. Following Ref. \cite{us2}, the improved action for $\mu
\ne 0$ can thus be obtained by introducing functions $f_n(\mu a) = 1 + n \mu a$
($g_n(\mu a) = 1 - n \mu a$) for $n$ =1, 2, and 3, which multiply the terms
containing forward (backward) links in the time direction. The coefficients in
eq. (\ref{baryon}) dictate that $f_n$ (or $g_n$) multiplies an $n$ time-link
term.

Without giving the details here, we assert that a) the improved action with
these $f_n$ and $g_n$ does lead in general to $\mu^2/a^2$ ($\mu/a^2$)
divergences in the energy (number) density and b) the ansatz $ f_n = f^n_1$,
$g_n = g^n_1$ with $f_1(\mu a) \cdot g_1(\mu a) =1$, eliminates them.  This
ansatz is a natural generalization of the results in \cite{us2} for the naive
action. Thus, as in the naive fermion case \cite{us1},  $f_1(\mu a)= \exp(\mu
a)$ or $f_1(\mu a) = (1 +\mu a)/\sqrt{1 - \mu^2 a^2}$ can be used for the
improved action in conjunction with the above ansatz for $f_n$ and $g_n$.

\section{NUMERICAL RESULTS}

Figure \ref{chiT} compares the behaviour of the ideal gas susceptibility at
$\mu =0 $, $\chi_{ideal}$, for the Naik and the P4 action with that for the
naive action.  The results have been obtained on lattices with an aspect ratio
of 3 ($N_s = 3 N_t$) and for a fixed quark mass $m/T$. Since $f'(0)= -g'(0)=
f''(0) = g''(0)= 1$ for all the allowed $f$ and $g$, the results in Figure
\ref{chiT} are valid for all prescriptions of $f$.  One sees that both the
improved actions result in a much milder $N_t$-dependence in $\chi_{ideal}$ and
the results are close to continuum by $N_t = 8$ already.  Interestingly, the
improved action results approach the continuum result from the opposite
direction as compared to the naive action.

\begin{figure}[htbp]\begin{center}
\vspace{-0.5cm}
\epsfig{height=6cm,file=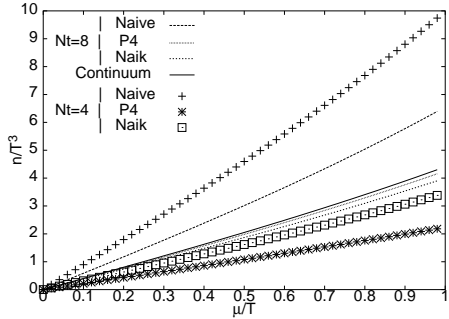, angle=270 }
\vspace{-0.7cm}
\caption{Comparison of the continuum number density as a function of $\mu/T$ 
with the naive, Naik and P4 improved actions on $12^3 \times 4$ and 
$24^3 \times 8$ lattices.} 
\vspace{-0.6cm}
\label{fg.nmt}\end{center}\end{figure}

As one switches on $\mu$, the susceptibilities for the improved actions
continue to be close to unity on lattices with $N_t \ge 8$ but they do exhibit
a mild $\mu$-dependence, increasing above unity for $\mu \sim T$. For small
$\mu$ one also sees a similar pattern in the number density, $n/T^3$, as seen
in Figure  \ref{fg.nmt}: the improved action results are closer to the continuum
results than the naive action for $N_t=8$.  The results also show a mild
$\mu$-dependence which may be a cause of concern for the methods which attempt
to study the equation of state at $\mu$ =0 by using a Taylor expansion in $\mu$
in the simulations and could result in systematic effects in the predictions
for the plasma phase.

\vspace{-0.3cm}

\begin{figure}[htbp]\begin{center}
\vspace{-0.5cm}
\epsfig{height=6cm,file=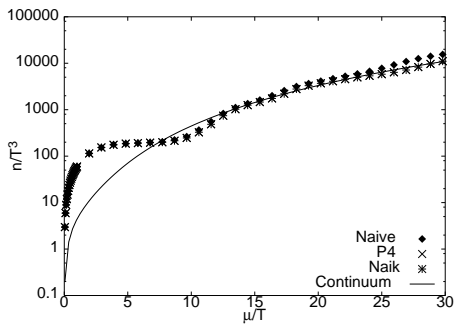, angle=270 }
\vspace{-0.7cm}
\caption{Same as Fig. \ref{fg.nmt} but on $24^3 \times 48$ lattices.}
\vspace{-0.6cm}
\label{fg.nmz}\end{center}\end{figure}
\vspace{-0.3cm}

Figure \ref{fg.nmz} shows the results obtained for the number density of an ideal
gas at very small temperatures by taking $N_t$ very large.  On a $24^3 \times
48$ lattice, the number density seems to be essentially the same for small
$\mu$ for all actions, including the naive action.  However, none of them has
the correct continuum $\mu$-dependence. Only when $\mu/T$ is large do they
all approach the continuum.  Improved actions seem to be closer to the
continuum result for large $\mu$.  Note that even at the highest $\mu/T$, 
$\mu a$ is still only about 0.6 in Fig. \ref{fg.nmz}.   The deviations
for small $\mu/T$ are due to the smallness of the spatial volume.  

\section{SUMMARY}

We have constructed an appropriate number density term, $N$, for a set of
improved actions from the current conservation equations for these actions.  The
Naik action and the P4 action, which correspond to specific choices of values
for the action parameters, are included in this set.  Adding $\mu N$ to the
action for $\mu=0$, where $\mu$ is the baryon chemical potential, leads to
divergences in the energy density and the number density, as for the unimproved
naive action.  These can be eliminated by the same condition as for the naive
action, namely, $f(\mu) \cdot g(\mu) = 1$, provided for each forward
(backward) time link  the corresponding gauge variable $U_0(x)$
($U^\dagger_0(x)$) is multiplied by $f(\mu)$ ($g(\mu)$).

The quark number susceptibility and the number density for the ideal gas do
approach the continuum result on smaller temporal lattices for the improved
action.  It would be interesting to study them in the high temperature regime
of QCD.  Finite size effects appear to be mildly $\mu$-dependent for the
susceptibility and the number density which may lead to systematic effects.

\end{document}